\documentclass[aps,pra,reprint,10pt,a4paper]{revtex4-1}
\usepackage[utf8]{inputenc}
\usepackage[sc,osf]{mathpazo}
\usepackage[T1]{fontenc}
\usepackage{amsfonts}
\usepackage{amssymb}
\usepackage{amsmath}
\usepackage{amsthm}
\usepackage{graphics}
\usepackage{graphicx}
\usepackage{braket}
\usepackage{xcolor}
\usepackage[colorlinks=true,linkcolor=blue,urlcolor=blue,citecolor=blue]{hyperref}

\begin{document}

\title{Detection of unbroken phase of non-Hermitian system   via Hermitian factorization surface} 
%
\author{Leela Ganesh Chandra Lakkaraju and Aditi Sen(De)}
 \affiliation{Quantum Information and Computation Group,
Harish-Chandra Research Institute,  A CI of Homi Bhabha National
Institute, Chhatnag Road, Jhunsi, Prayagraj - 211019, India}

\begin{abstract}
In the traditional quantum theory, one-dimensional quantum spin models possess a factorization surface where the ground states are fully separable having vanishing bipartite as well as multipartite entanglement.  We report that in the non-Hermitian counterpart of these models, these factorization surfaces either can  predict  the exceptional points where the unbroken-to-broken transition occurs or can guarantee the reality of the spectrum, thereby proposing a procedure  to reveal the unbroken phase. We first analytically demonstrate it  for the nearest-neighbor rotation-time (\(\mathcal{RT}\))-symmetric $XY$ model with uniform and alternating transverse magnetic fields, referred to as the \(iATXY\) model. Exact diagonalization techniques are then employed to establish this fact for the \(\mathcal{RT}\)-symmetric $XYZ$ model with short- and long-range interactions as well as for the long-ranged \(iATXY\) model.
Moreover, we show that although the factorization surface prescribes the unbroken phase of the non-Hermitian model, the bipartite nearest-neighbor entanglement at the exceptional point is nonvanishing.

\end{abstract}

\maketitle

\section{Introduction}
\label{sec:intro}

Over the years, studying the phenomena and  properties of one-dimensional short-ranged quantum spin models in the presence of magnetic fields  has developed lots of interest \cite{qptbook2, qptbook1} since several such Hamiltonians can be mapped  to spinless fermions \cite{spinless} and hard-core bosons \cite{hardboson}, thereby  ensuring the analytical study of  single-, two- and multi-site features. Moreover,  they can be simulated in laboratories with physical substrates like ultracold atoms \cite{ultracold-review}, nuclear magnetic resonances \cite{nmr_book,nmr_aditi,nmr1} and ion traps \cite{iontrapBlatt}. Apart from investigating phenomena like quantum phase transitions at zero-temperature and the quantum dynamical transition in evolution,  these systems have been shown to be potential candidates for designing quantum technologies \cite{oneway1, oneway2, oneway3, quantumbattery_review}. Moreover,  these models also possess a factorization surface or volume in the parameter space, \cite{Fact1, XYZfac, fac-ger, facrev1, facrev3, facrev4, facrev5} at which the ground state  is doubly degenerate and is fully separable, having vanishing bipartite as well as multipartite entanglement \cite{ent1}.  

On the other hand, in the seminal paper by Bender and  Boettcher \cite{Bender'98}, it was shown that 
non-Hermitian Hamiltonian with both parity and time reversal symmetry, (together called \(\mathcal{PT}\)-symmetry), can have real energy spectrum while the breaking of symmetry leads to the complex eigenenergy. The phase transition from symmetry broken to an unbroken phase occurs at the exceptional point. These results  simulate a significant amount of research to characterize non-Hermitian quantum theory, both theoretically and experimentally, especially in optics \cite{ptsymmetry_optics}, cold atoms \cite{ptsymmetry_open}, cavities \cite{ptsymmetry_microcavity, ptsymmetry_atom_cavity}. In this respect, discrete systems like tight binding model, quantum spin systems, specifically, one-dimensional quantum $XY$ models  turn out to be  important platforms to verify the properties in non-Hermitian Hamiltonian \cite{PTSymmetry_tightbinding_1, PTSymmetry_tighbinding_2,  PTSymmetry_tighbinding_4, PTSymmetry_spin1, PTSymmetry_spin2, PTSymmetry_statetransfer, PTSymmetry_network, PTSymmetry_bosehubbard, PTSymmetry_scattering}. It was also noticed that instead of \(\mathcal{PT}\)-symmetry, linear rotation operator, \(\mathcal{R}\), which rotates each spin by a certain amount around a fixed axis, along with time reversal operator \(\mathcal{T}\)  can together prompt  non-Hermiticity  in the quantum spin models \cite{SongReal}. Specifically, it was shown that the nearest-neighbor transverse $XY$ model with imaginary anisotropy  parameter has  \(\mathcal{RT}\)-symmetry and  undergoes a transition from the unbroken phase to a broken one which can again be detected via the existence of change in the spectrum from real to imaginary ones.  
In both  non-Hermitian fermionic and bosonic systems \cite{RTSymmetry_boson}, Berry curvature, \cite{RTSymmetry_berrycurvature} and multipartite entanglement \cite{RTSymmetry_entanglement} are used to describe the broken-to-unbroken transitions.

In the current work, we first investigate the one-dimensional (1D) \(\mathcal{RT}\)-symmetric nearest-neighbor $XY$ model in presence of uniform as well as alternating magnetic fields \cite{atxy} which we call  as the \(iATXY\) model. 
The Hermitian version of this model possesses a richer phase diagram than that of the transverse $XY$ model. In particular, it has paramagnetic-II (PM-II) phase with a large amount of bipartite entanglement along with antiferromagnetic (ferromagnetic), and  paramagnetic-I phases \cite{ATXY_phases_amit,ATXY_phase}. Moreover,  like the $XY$ model, it can also be diagonalized by  Jordan-Wigner, Fourier, and Bogoliubov transformations \cite{LSM,SongReal,jordan-wigner,Ising_exact,qptbook2}. By employing similar transformations in the non-Hermitian case, we report that the exceptional points which divide the real and imaginary spectrum can be inferred from the factorization surface of the corresponding Hermitian Hamiltonian. The finite-size exact diagonalization calculations also confirm this result, thereby motivating us to consider  quantum spin models which cannot be solved analytically.

To establish the relation between the broken to the unbroken transition of the non-Hermitian model and the factorization surface of the Hermitian counterpart, we 
study both nearest-neighbor and long-ranged \(\mathcal{RT}\)-symmetric  $XYZ$ model as well as \(iATXY\) model with long-range interactions. In all these systems, numerical simulations strongly suggest that the unbroken phase of the \(\mathcal{RT}\)-symmetric models can be identified by the corresponding Hermitian  factorization surface. Specifically, we find that the energy spectrum is real at and above the surface predicted via the factorization surface of the  Hermitian Hamiltonian, thereby providing a sufficient condition for the reality of the spectrum. Interestingly, we observe that at the surface, the bipartite nearest neighbor entanglement of the  \(iATXY\) model is nonvanishing. At this point, we are tempted to  conjecture that tuning parameter of  the \(\mathcal{RT}\)-symmetric Hamiltonian which leads to a real spectrum can be determined from the factorization surface of the corresponding Hermitian models.
It is important for the following three reasons. (1)  In  both the Hermitian and non-Hermitian domains, there are quantum spin models,  for which the spectrum can only be found numerically although the factorization surfaces are known from different symmetry properties of the system \cite{fac-ger}.
(2) Finding where the spectrum is real is not an easy task, owing to the computational problems of diagonalizing non-Hermitian Hamiltonians although we know that any physics that is observable and measurable needs to be done when the spectrum is real. Our method directly prescribes either an exceptional point for the nearest-neighbor model or suggests parameters that correspond to the real spectrum, thereby simplifying the situation enormously.
(3) Lastly, our results possibly show that the properties of the Hermitian system  have the potential to diagnose the exceptional point of the non-Hermitian systems without computation.

The paper is organized as follows. In Sec.  \ref{sec:ATXY}, we discuss the way to diagonalize the pseudo-Hermitian \(ATXY\) model while the broken to unbroken transition identified via factorization surface of the corresponding Hermitian model is presented in Sec. \ref{sec:transition}. In the Sec. \ref{sec:numerical}. we confirm that the exceptional point is predicted by the factorization surface by considering nearest-neighbor  \(\mathcal{RT}\)-symmetric \(XYZ\) model. In SubSec. \ref{subsec:LR},  the procedure for detecting the unbroken phase in both the models with long-range interactions having RT symmetry  is provided.The behavior of bipartite entanglement and parity in the transition surface of the \(iATXY\) model are described in Sec. \ref{sec:ent} and we conclude in Sec. \ref{sec:conclu}.


\section{Pseudo-Hermitian $iATXY$ model}
\label{sec:ATXY}

Let us first consider the pesudo-Hermitian  one-dimensional nearest-neighbor $XY$ model with imaginary anisotropy factor 
in the presence of uniform and alternating transverse magnetic field. The Hamiltonian reads as,
\begin{eqnarray}
    \hat{H}_{iATXY} &=& \sum_{i=1}^{N} J \Big[ \frac{1+i\gamma}{4}\sigma_i^x\sigma_{i+1}^x + \frac{1-i\gamma}{4}\sigma_i^y\sigma_{i+1}^y \Big ] \nonumber \\
&+& \frac{h_1 + (-1)^i h_2}{2} \sigma_i^z,
\label{eq:ham1}
\end{eqnarray}
where $J \ne 0$ is the coupling constant,  $\sigma^{x,y,z}$ are Pauli matrices,  and $i\gamma$ is the imaginary anisotropy parameter while  $h_1 - h_2$ and \(h_1 + h_2\) are the strengths of magnetic fields on odd and even spins respectively. We also assume periodic boundary condition throughout the paper, i.e., \(\sigma_{N+1} = \sigma_1\). 
Like the $XY$ model with uniform field \cite{SongReal}, we will show that the non-Hermitian \(\hat{H}_{iATXY}\) with  $\mathcal{RT}$ symmetry has real spectrum in the unbroken phase, while the complex eigenenergy is found in the broken phase. Here   $\mathcal{R}$ is the application of $\frac{\pi}{2}$ rotation about the $z$ axis, given by
\(\mathcal{R} \equiv e ^{\left[-i(\pi / 4) \sum_{j=1}^{N} \sigma_{j}^{z}\right]}\),
and the time reversal, $\mathcal{T}$, is the complex conjugation in case of finite dimensional systems. The Hamiltonian is not individually invariant under either $\mathcal{R}$ or $\mathcal{T}$ operators  but when combined, the Hamiltonian is invariant under  $\mathcal{RT}$, i.e.,  $[H,\mathcal{RT}] = 0$.
As shown in Ref. \cite{SongReal}, the  effects of $\mathcal{RT}$ symmetry are similar to that of  $\mathcal{PT}$ symmetry. In particular, 
the Hamiltonian always commutes with $\mathcal{R}T$, although $H$ and $\mathcal{RT}$ do not always share the same eigenvectors due to the anti-linearity of $\mathcal{T}$, which leads to the parametric dependence  having a real spectrum.

 \subsection{Energy spectrum of \(iATXY\) model}
\label{sec:diag}

 By performing Jordan-Wigner transformation followed by a Fourier transform of the fermionic operators, the $iATXY$ model in Eq. (\ref{eq:ham1}) can be diagonalized by employing the similar procedure as Hermitian $ATXY$ model \cite{atxy,ATXYDiag1,ATXYDiag2,ATXYDiag3,ATXY_phases_amit,ATXY_phase}. First, let us convert $H_{iATXY}$ in terms of  $\sigma^+$ and  $\sigma^-$  operators where $\sigma^x = \frac{\sigma^++\sigma^-}{2}$, $\sigma^y = \frac{\sigma^x-\sigma^y}{2i}$ and $\sigma^z = \sigma^+\sigma^- - \frac{1}{2}$. 
The Jordan-Wigner transformation
\begin{equation*}
\begin{array}{l}
\hat{\sigma_{2j} }^{+}=\hat{e}_{2 j}^{\dagger} \exp \left(i \pi \sum_{l=1}^{i-1} \hat{e}_{2 l}^{\dagger} \hat{e}_{2 l}+i \pi \sum_{l=1}^{i} \hat{o}_{2 l-1}^{\dagger} \hat{o}_{2 l-1}\right) \\
\hat{\sigma}_{2 j+1}^{+}=\hat{o}_{2 j+1}^{\dagger} \exp \left(i \pi \sum_{l=1}^{i} \hat{e}_{2 l}^{\dagger} \hat{e}_{2 l}+i \pi \sum_{l=0}^{i-1} \hat{o}_{2 l+1}^{\dagger} \hat{o}_{2 l+1}\right)
\end{array}
\end{equation*} 
 maps  the system into a 1D two-component Fermi gas, where the even and odd sites correspond to fermions, following the fermionic commutation rules, governed by $\hat{e}$ and  $\hat{o}$ respectively.

The parity operator  defined as, $\xi=\prod_{i=1}^{N}\left(\sigma_{i}^{z}\right)=(-1)^{N_{f}}$, where $N_f = \sum_{i=1}^{\frac{N}{2} } \hat{o}_{2 i-1}^{\dagger} \hat{o}_{2 i-1} + \hat{e}_{2 i}^{\dagger} \hat{e}_{2 i} $ commutes with the Hamiltonian, i.e., $[H,\xi]=0$ which is the sum of the number of fermions.  Considering the parity and the Jordan-Wigner transformation, the Hamiltonian can be written as
\begin{equation}
\begin{aligned}
\hat{H}_{iATXY}=& \sum_{i=1}^{\frac{N}{2} -1 }\left[\left\{\hat{\mathcal{X}}_{i}+\hat{\mathcal{Y}}_{i}+i\gamma\left(\hat{\mathcal{V}}_{i}+\hat{\mathcal{W}}_{i}\right)\right\}\right.\\
&\left.+h_{o}\hat{\mathcal{M}}_{i}^{o}+h_{e} \hat{\mathcal{M}}_{i}^{e}\right]-\mu(\hat{\mathcal{Y}}_{\frac{N}{2}}+ i\gamma\hat{\mathcal{W}}_{\frac{N}{2}}),
\end{aligned}
\end{equation}
where $\hat{\mathcal{X}}_{i}=\hat{o}_{2 i-1}^{\dagger} \hat{e}_{2 i}+$ H.c,  $\hat{\mathcal{Y}}_{i}=\hat{e}_{2 i}\hat{o}_{2 i+1}^{\dagger} +$ H.c, $\hat{\mathcal{V}}_{i}=\hat{o}_{2 i-1}^{\dagger} \hat{e}_{2 i}^{\dagger}+$ H.c.,  $\hat{\mathcal{W}}_{i}=\hat{e}_{2 i}^{\dagger}\hat{o}_{2 i+1}^{\dagger} +$ H.c and the last term is for the boundary condition with \(\mu\) being the eigenvalues of \(\xi\) with distinct values \(\pm 1\). Here the number of odd and even fermions are given by the $\hat{\mathcal{M}}_{i}^{o} = \hat{o}_{2 i-1}^{\dagger} \hat{o}_{2 i-1} $ with the field $\lambda_o = 2(h_1 - h_2)/J$ and  $\hat{\mathcal{M}}_{i}^{e} = \hat{e}_{2 i}^{\dagger} \hat{e}_{2 i} $ with the field $\lambda_e = 2(h_1 +h_2)/J$. We set \(h_i/J = \lambda_i\), \(i=1, 2\). 
Using Fourier transformations, given by
\begin{equation}
\begin{aligned}
\hat{o}_{2 j+1}^{\dagger} &=\sqrt{\frac{2}{N}} \sum_{p=-N / 4}^{N / 4} \exp \left[i(2 j+1) \phi_{p}\right] \hat{o}_{p}^{\dagger}, \\
\hat{e}_{2 j}^{\dagger} &=\sqrt{\frac{2}{N}} \sum_{p=-N / 4}^{N / 4} \exp \left[i(2 j) \phi_{p}\right] \hat{e}_{p}^{\dagger},
\end{aligned}
\end{equation}
we write the Hamiltonian including the boundary terms, i.e.,  the  summation of $\sum_{i=1}^\frac{N}{2}$ which can be done by considering suitable Fourier momenta: odd parity ($\mu = -1$) is given by $k^- =\frac{2\pi p}{N}$ and even parity ($\mu = +1$) is given by $k^+ =\frac{2\pi(p+1/2)}{N}$. The Hamiltonian can  now be written in the Fourier basis,  $S^k = \{o_{k^\mu}, o^\dagger_{-{k^\mu}},e_{k^\mu}^\dagger, e_{-{k^\mu}}\} $, as 
\begin{equation}
H_{iATXY} =\sum_{k \in {k^\mu}} H^{{k}}_{iATXY}=\sum_{k \in {k^\mu}} (\hat{S}^{k})^{\dagger} \hat{H}_{iATXY}^{{k}} \hat{S}^{{k}}.
\end{equation}
Since the Hamiltonian is invariant under parity, the corresponding $k^\pm$  do not mix, and hence we can consider, $k$ as a general momentum running through both even and odd momenta. Thus, the Hamiltonian  $H^k_{iATXY}$ reduces to 
\begin{equation}
\left[\begin{array}{cccc}
\lambda_1+ \cos k & - \gamma \sin k & 0 & -\lambda_2 \\
 \gamma \sin k & -\lambda_1- \cos k & \lambda_2 & 0 \\
0 & \lambda_2 &  \cos k-\lambda_1 & - \gamma \sin k \\
-\lambda_2 & 0 &  \gamma \sin k & - \cos k+\lambda_1
\end{array}\right].
\end{equation}
Diagonalizing $H^k_{iATXY}$ gives the single-particle energy spectrum of the model in each $k$ subspace in terms of \(\lambda_i\), (\(i=1, 2\)) and \(\gamma\) as
\begin{equation}
\begin{aligned}
E^{k}_{\pm}=&\left[\lambda_1^{2}+\lambda_2^{2}+ \cos ^{2} k\right.\\
&\left.-\gamma^{2} \sin ^{2} k \pm 2 \sqrt{\lambda_1^{2} \lambda_2^{2}+h_1^{2} \cos ^{2} k+ \lambda_2^{2} \gamma^{2} \sin ^{2} k}\right]^{1 / 2},  
\end{aligned}
\label{eq:energy}
\end{equation}
which finally leads to the energy spectrum  of the model and hence can be used to obtain the exceptional points, dividing the regions of broken and unbroken phases in the \(iATXY\) model.  

\section{Broken-unbroken transition of the quantum $iATXY$ model at the factorization surface of the Hermitian   model }
\label{sec:transition}


Having obtained the energy for each momentum space, \(k\), let us concentrate on the transition from the broken to unbroken phase. In other words, in the unbroken phase, 
the spectrum becomes real if the same set of eigenvectors spans both $H$ and $\mathcal{RT}$, while in the broken phase, complex conjugate pairs are the eigenvalues of \(\hat{H}_{iATXY}\).
 \begin{figure*}
    \centering
    \includegraphics[scale = 0.4]{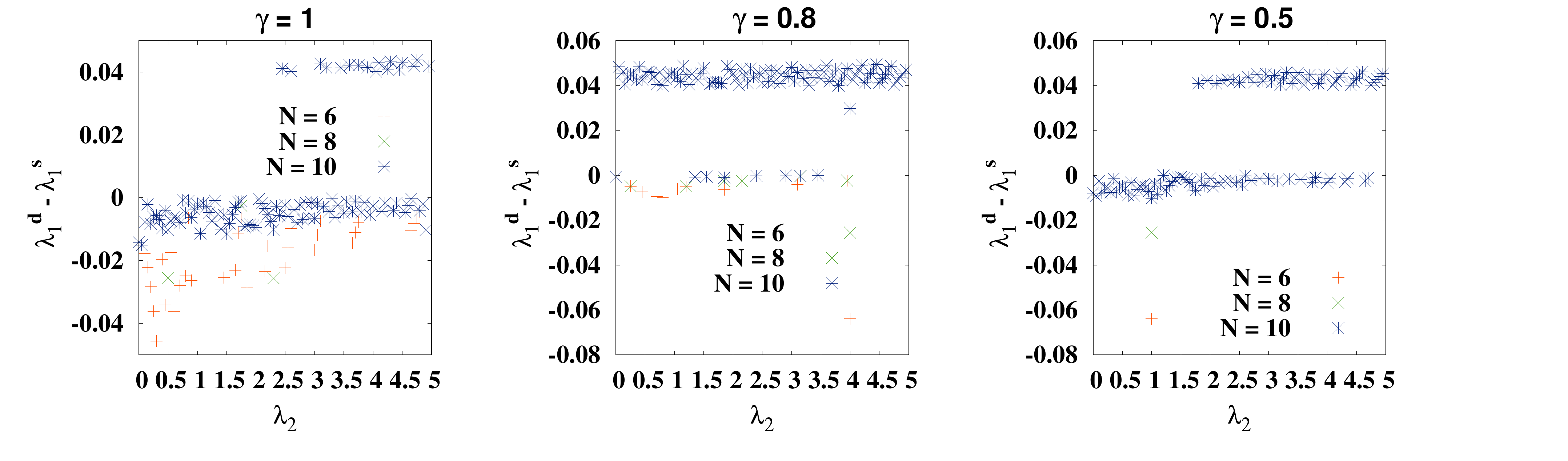}
    \caption{Detected numerical value vs. prediction of the \(iATXY\) model. The difference between the detected value, $\lambda_1^d$ found numerically  (the value  of \(\lambda_1\) at which the spectrum becomes real) and the predicted value, $\lambda_1^s$  (according to  (\ref{realEq})) (vertical axis) against $\lambda_2$ (horizontal axis). The anisotropy of \(\hat{H}_{iATXY}\), \(\gamma\), are fixed in each plot as mentioned at the top. The difference decreases  with the increase of system-sizes ($N = 6,8$ and $10$ are chosen to show convergence). Both the axes are dimensionless.}
    \label{fig:ATXY_1}
\end{figure*}
Let us identify the parameter space where the spectrum is real. To identify it, we will be looking for the value of $k$ at which $E^{k}_{\pm}$ has an extremum and   $(E^{k}_{\pm})^2 \geq 0$. It turns out that 
the value of $k$ for which $E^{k}_{\pm}$ reaches its extremum, i.e.,  \(\frac{d E^k_{\pm}}{d k} =0\), is given by
\begin{equation}
    k = \cos^{-1}\left(\sqrt{\frac{1}{\lambda_1^2 + \lambda_2^2\gamma^2}\Big[\frac{\lambda_1^2 + \lambda_2^2 \gamma^2}{1+\gamma^2}\Big] + \lambda_2^2\gamma^2 - \lambda_1^2 \lambda_2^2 }\right).
\end{equation}
Plugging it into  $E^k_-$ and checking when  it is real, which is equivalent to find out when $(E^k_-)^2 \ge 0$, we find that the parameter space is split  in order to have real spectrum when
\begin{equation}
    \begin{aligned}
     \lambda_1 & \geq \sqrt{1+\lambda_2^2+\gamma^2} \equiv \lambda_1^s , \text{   }  \text{   } \lambda_1 > \lambda_2, \\
     \lambda_1 & \leq \sqrt{1+\lambda_2^2+\gamma^2} , \text{   } \text{   }  \lambda_1 < \lambda_2.
    \end{aligned}
    \label{realEq}
\end{equation}
Let us first note that the second case is not possible. The reason is that  when $\lambda_1 < \lambda_2$, i.e., when $\lambda_1^2 - \lambda_2^2 < 0$, the real eigenvalues occur  
for $\lambda_1^2 -\lambda_2^2 > 1+\gamma^2 $ which is not possible since $\gamma$ is real. Hence the transition from the broken to unbroken phase happens when \( \lambda_1 \geq \sqrt{1+\lambda_2^2+\gamma^2}\). Notice that for the uniform $XY$ model, i.e., with \(\lambda_2=0\), the eigenvalues go from real to complex pairs when \(\lambda_1 \geq \sqrt{1 + \gamma^2}\) as found in Ref. \cite{SongReal}. Notice also that \(E^k+\) does not lead to any useful condition in terms of \(\lambda_1,\lambda_2\) and \(\gamma\).


Apart from quantum critical points,  the Hermitian \(ATXY\) model possesses a special point (surface) known as
the factorization surface \cite{ATXY_phase, atxy, Fact1},  denoted by  \(\lambda_{1(H)}^f\), which can be represented as 
\begin{equation}
     \lambda_{1(H)}^f = \sqrt{1+\lambda_2^2-\gamma^2}
     \label{eq:ATXYfac}
\end{equation}
in the parameter space. At this surface,  the ground state  is doubly degenerate and fully separable, having vanishing bipartite as well as multipartite entanglement. 

If we now replace \(\gamma\) by \(i \gamma\) in Eq. (\ref{eq:ATXYfac}), we recover the first condition of having real spectrum of the \(iATXY\) model given  in  (\ref{realEq}). We denote the right hand side of  (\ref{realEq}) as \(\lambda_1^s\). This suggests that the transiton from the symmetry-broken phase to the unbroken phase of the \(\mathcal{RT}\)-symmetric Hamiltonian can be identified by the factorization surface of the corresponding Hermitian Hamiltonian. 

Therefore, we propose the following:  if the Hermitian Hamiltonian has a factorization surface, \(\Lambda^f_{(H)} (\eta, \eta', \ldots)\), which is specified by the parameters of the Hamiltonian, \(\eta, \eta', \ldots\), the corresponding \(\mathcal{RT}\)-symmetric Hamiltonian possess real eigenvalues when  \(\Lambda \geq \Lambda^s (i\eta, i\eta', \ldots)\) where some parameters can be complex to  preserve the symmetry of the Hamiltonian.


Since the $iATXY$ model can be solved analytically,  we are able to derive the transition surface analytically. The above interesting observation can give us an important tool for detecting the phases of the non-Hermitian models, especially those models which cannot be solved analytically. \\

Remark. It is important to notice that the property obtained above is exclusive to  the $\mathcal{RT}$-symmetric non-hermitian system as opposed to the $\mathcal{PT}$ symmetric ones.  For example,  it was reported that a dimerized spin system with added imaginary local magnetic field with strength $\eta$ conforms to $\mathcal{PT}$ -symmetriy, having a real spectrum only when $\eta < \eta_c$ where the $\eta_c=\min  [ \mbox{sum of the interaction strength in x and y directions}\), \( \mbox{difference of interaction strength in x and y directions}]$ \cite{PTSymmetry_spin2}, which is not dictated by the factorization point \cite{dimefact}.


In the rest of the paper, we demonstrate that the known factorization surface of the Hermitian model can indeed give the sufficient condition for the non-Hermitian nearest-neighbor $iXYZ$ model with imaginary $\gamma$, as well as for the fully connected $iATXY$  and the $iXYZ$ models. 
Before addressing these models which cannot be solved analytically, we will now check whether the condition for real eigenvalues of the \(iATXY\) model in  (\ref{realEq}) match the numerically obtained condition for real energies. Specifically,   
for a fixed \(N\),  $\lambda_2$ and \(\gamma\),  we  search numerically for $\lambda_1$,  which 
gives the entire spectrum as real and we match the detected value, \(\lambda_1^d\) with  $\lambda_1^s$  obtained from the condition (\ref{realEq}). 

We use exact diagonalization technique which utilizes Krylov subspace method, known as Lanczos method \cite{Lanczos}. Although it was noted  \cite{PTSymmetry_numerical_arnoldi} that  Arnoldi method may give better results for \(\mathcal{PT}\)-symmetric systems, we observe that there is no  qualitative difference between Lanczos and Arnoldi \cite{arnoldi} methods. Both of these numerical mechanisms are part of ARMADILLO package \cite{armadillo1,armadillo2} which we use to analyze our systems.

For a fixed anisotropy, \(\gamma\), Fig. \ref{fig:ATXY_1} depicts the behavior of the difference, \(\lambda_1^d - \lambda_1^s\) against \(\lambda_2\) from \(iATXY\) model for different system-sizes. 
Note that the numerical error is of the order of $\pm 0.05$, which is the same as the step size of $\lambda_2$. 
Figure shows that our inferrence is in accordance with the  numerical data under the specified numerical accuracy which increases with the increase in the system-size.


\begin{figure*}
    \centering
    \includegraphics[scale= 0.4]{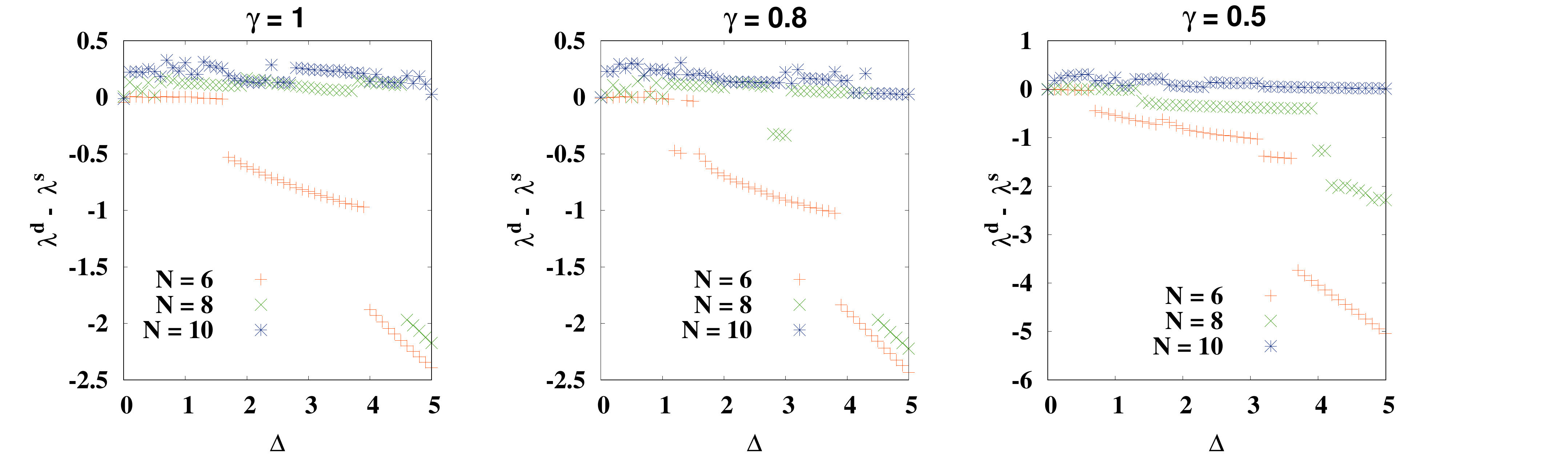}
    \caption{Plot of \(\lambda^d -\lambda^s\) (vertical axis)  vs.  $\Delta$ (horizontal axis) of the \(iXYZ\) Hamiltonian, \(H_{iXYZ}\). The similar analysis as in Fig. \ref{fig:ATXY_1} is performed for  the $iXYZ$ model.  Notice that  for low system-size, \(N\), the numerical values substantially divert from the inferred values for large \(\Delta\).    All other specifications are same as in Fig. \ref{fig:ATXY_1}.} 
    \label{fig:XYZ_1}
\end{figure*}

\section{Connecting the factorization point of the  Hermitian model with the unbroken phase of \(\mathcal{RT}\)-symmetric models: short- to long-range interactions}
\label{sec:numerical}

In this section, we consider nearest-neighbor as well as long-ranged $XYZ$ models with magnetic field having imaginary anisotropy parameter and also long-ranged  \(iATXY\) models. All these models have \(\mathcal{RT}\) symmetry although they cannot be solved analytically. We apply exact diagonalization tool, as mentioned in the previous section to diagonalize the pseudo-Hermitian Hamiltonian and find the parameter space in which the eigenvalues are real. 

\subsection{$iXYZ$ model: Numerical vs. prediction}

Let us first consider the non-Hermitian nearest-neighbor $XYZ$ Hamiltonian with  $\mathcal{RT}$-symmetry, which we call as $iXYZ$ model, given by
\begin{equation}
\begin{aligned}
 \hat{H}_{iXYZ} &=  \sum_{i=1}^N J \Big [ \frac{1+i\gamma}{4}\sigma_i^x\sigma_{i+1}^x + \frac{1-i\gamma}{4}\sigma_i^y\sigma_{i+1}^y + \frac{\tilde{\Delta}}{4}\sigma_i^z\sigma_{i+1}^z\Big]\\
& + \frac{h}{2} \sigma_i^z,
\end{aligned}
\label{eq:ham2}
\end{equation}

where \(\tilde{\Delta}\) is the strength of the  interaction in the \(z\)-plane and other parameters are same as in \(\hat{H}_{iATXY}\).  Here we set \(\Delta = \tilde{\Delta}/J\) and \(\lambda = h/J\). It can be easy to find that \( [\hat{H}_{iXYZ}, \mathcal{RT}]=0\). 
Since we cannot diagonalize this Hamiltonian analytically, let us follow the prescription mentioned above to find the parametric condition for which  the Hamiltonian has real eigenvalues.
In this case, the factorization surface of the Hermitian $XYZ$ model \cite{XYZfac} is known to be 
\begin{equation}
    \lambda^f_{(H)} = \sqrt{(1+\Delta)^2-\gamma^2}.  
\end{equation}
We propose that the spectrum becomes real when the magnetic field satisfies the condition given by 
\begin{equation}
    \lambda \ge \lambda^s \equiv \sqrt{(1+\Delta)^2+\gamma^2}. 
    \label{eq:faciXYZ}
\end{equation} 
For a given $\gamma$ and $\Delta$, we numerically find the actual $\lambda^d$ for which all eigenvalues are real. In Fig. \ref{fig:XYZ_1},  for three different values of $\gamma$, the difference between the detected magnetic value $\lambda^d$ and the predicted value, \(\lambda^s\), according to  (\ref{eq:faciXYZ}) is plotted.  We observe that with the increase of $N$, \((\lambda^d  -\lambda^s)\) are of the order of $\pm 0.05$ where  the step size of $\Delta$ is also considered to be $0.05$.  As shown in the case of the \(iATXY\) model, we can also report here that the prediction and the numerical values are in a good agreement, thereby verifying the prescription proposed to find the reality of the spectrum. 



\subsection{Pseudo-Hermitian model with long-range interactions}
\label{subsec:LR}

Up to now, all the spin models that we discussed have the nearest-neighbor interactions and we show that the unbroken to broken transition can faithfully be detected via the factorization surface of the respective Hermitian model. Let us now move to $iATXY$ as well as $iXYZ$ models having long-range interactions and exhibit whether the sufficient condition of identifying reality of the spectrum still remains valid or not.  
It is important to note here that the long ranged models are more natural to occur in experiments \cite{RevModPhyspower, Neyenhuise1700672, Islam583, CSB1, CSB2, lahaye2009physics, Jurcevic2014} and restricting  interactions to just nearest neighbors is a  tedious task in laboratories.  Hence a more experimental-friendly model is the one where the strength of the interactions decreases as the distance between the neighbors increases. We now carry out our analysis with this kind of  models having  $\mathcal{RT}$ symmetry. In order to build the long ranged model with $\mathcal{RT}$ symmetry, we realize that other than the anisotropy strength involved in  the interactions of the \(xy\)-plane, we should not add imaginary terms in $\Delta$ or $\lambda_2$ since they fail to keep the symmetry.

\subsubsection{$iATXY$ model with long-range interactions}

\begin{figure*}
    \centering
    \includegraphics[width = \linewidth]{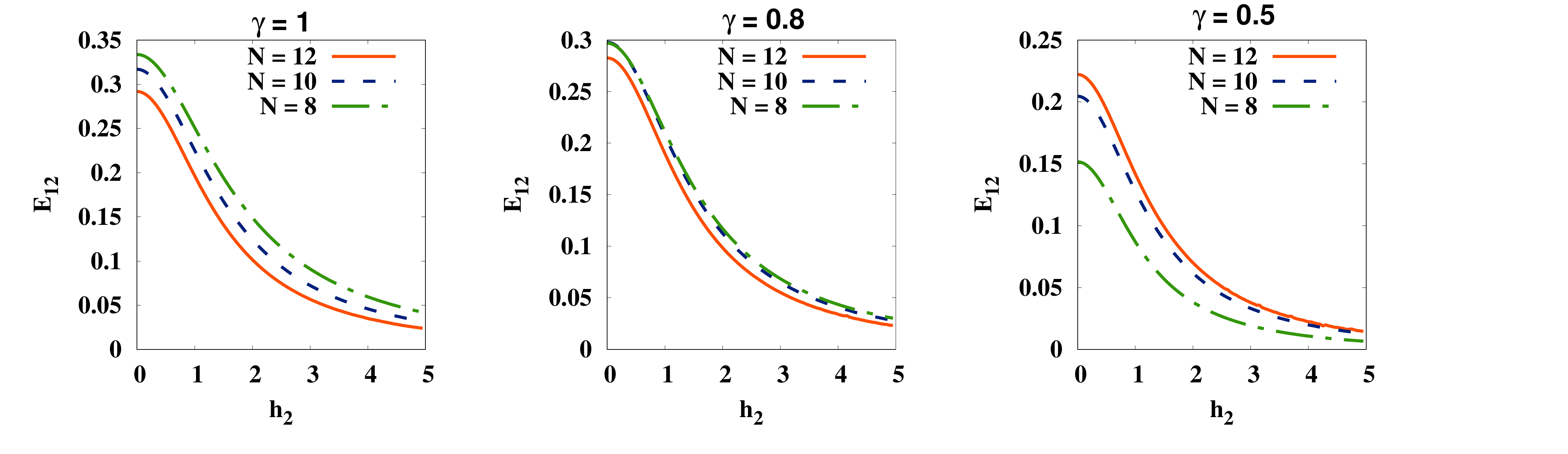}
    \caption{Nearest neighbor entanglement, \(E_{12}\) (ordinate),  of the zero-temperature state of the \(iATXY\) model at the exceptional point  against \(\lambda_2\) (abscissa) for different system sizes. Here \(N =8\) (long-short dashed lines), \(N=10\) (dashed lines) and \(N=12\) (solid lines). 
    For a given $\gamma$ and $\lambda_2$ values,  $\lambda_1^s$ is obtained via the condition \ref{realEq}. After tracing out every spin except first and second, we find $\rho_{12}$ and compute bipartite entanglement measured by logarithmic negativity  ($E_{12})$ \cite{neg4, log_neg}. Both the axes are dimensionless.  }
    \label{fig:NNent}
\end{figure*}

Consider the $iATXY$ model with long-range interactions, given by
\begin{equation}
\begin{aligned}
 \hat{H}^L_{iATXY} = \sum_{i=1}^N \sum_{ j = i+1 }^{i+\frac{N}{2}} &J_{ij} \Big[ \frac{1+i\gamma}{4}\sigma_i^x\sigma_{j}^x + \frac{1-i\gamma}{4}\sigma_i^y\sigma_{j}^y \Big ] \\
&+ \frac{h_1 + (-1)^i h_2}{2} \sigma_i^z,
\end{aligned}
\label{eq:hamLR1}
\end{equation}
where the parameters except \(J_{ij}\) have the same  features as in \(H_{iATXY}\) in Eq. (\ref{eq:ham1}). 
Here we consider power-law interactions, i.e., $J_{ij} = \frac{J}{|i-j|^\alpha}$, where $\alpha$ dictates how fast the interaction falls off with distance. 
For example,  a very high $\alpha$ value essentially imitates a nearest neighbor model while a low value corresponds to the situation when all of the spins are interacting with every other spin.

 In this case, the factorization surface   \cite{facrev1, facrev2, facrev3} is given to be
\begin{equation}
    \lambda_{1(H)}^{f}(\alpha) = \sqrt{1+\lambda_2^2-\gamma^2} \sum_{j=i+1}^{i+\frac{N}{2}} \frac{1}{|i-j|^\alpha},
\end{equation}
and hence according to our recipe,  the  spectrum of \(\hat{H}^L_{iATXY}\) is real when  
\begin{equation}
   \lambda_1^\alpha \geq \lambda_1^{s}(\alpha) \equiv 
   \sqrt{1+\lambda_2^2+\gamma^2} \sum_{j=i+1}^{i+\frac{N}{2}} \frac{1}{|i-j|^\alpha}.
\end{equation}
By performing exact diagonalization of $\hat{H}^L_{iATXY}$ for different system sizes, we uncover that  for a fixed \(\lambda_2^\alpha\), the difference between our prediction and the value of $\lambda_1^\alpha$ at which the entire spectrum becomes real is not exactly zero.  The reason behind such an observation is that the spectrum starts becoming real for some range of $\lambda_1^\alpha$ and then again becomes imaginary, thereby  creating a difficult situation  for finding the exact transition point. However,  when we start looking at and above  $\lambda_1^{s}(\alpha = 1)$, we find that the eigenvalues always remain real. To ensure that this is true,  in steps of $0.05$, we check from the predicted $\lambda_1^{s}(\alpha = 1)$ to $\lambda_1^{s}(\alpha = 1) + 5.0$ and confirm that at all of $100$ points, the spectrum is real for a given \(\lambda_2^\alpha\). Thus,  as prescribed, \( \lambda_{1(H)}^{f}(\alpha)\) of the Hermitian model  can suitably predict \(\lambda_1^s\) which provides a sufficient condition for  the unbroken phase of the pseudo-Hermitian model.

\subsubsection{Long-range $iXYZ$ model:  Sufficient condition for the unbroken phase } 

Let us now analyze the $\mathcal{RT}$-symmetric $iXYZ$ model when it is fully connected according to the power-law decay, represented as
\begin{equation}
\begin{aligned}
 \hat{H}_{iXYZ}^L = \sum_{i=1}^N \sum_{ j = i+1 }^{i+\frac{N}{2}} &J_{ij} \Big[ \frac{1+i\gamma}{4}\sigma_i^x\sigma_{j}^x + \frac{1-i\gamma}{4}\sigma_i^y\sigma_{j}^y + \frac{\Delta}{4}\sigma_i^z\sigma_j^z\Big]\\
&+ \frac{h}{2} \sigma_i^z, 
\end{aligned}
\label{eq:hamLR2}
\end{equation}
where \(J_{ij}\) behave similarly as in Eq. (\ref{eq:hamLR1}). 
The factorization surface of the corresponding Hermitian model reads as
\begin{equation}
    \lambda^{f}_{(H)}(\alpha) = \sqrt{(1+\Delta)^2-\gamma^2}  \sum_{j=i+1}^{i+\frac{N}{2}} \frac{1}{|i-j|^\alpha},
\end{equation}
which suggests that the point at which the eigenvalues of $\hat{H}_{iXYZ}^L$ become real is
\begin{equation}
   \lambda (\alpha) \geq  \lambda^{s}(\alpha) = \sqrt{(1+\Delta)^2+\gamma^2}  \sum_{j=i+1}^{i+\frac{N}{2}} \frac{1}{|i-j|^\alpha}.
\end{equation} 
Like in the long-range \(iATXY\) model,  for a given \(\Delta\), finding  \(\lambda(\alpha)\) at which the spectrum becomes completely real, is hard to find numerically. However, we apply the same method as before, i.e., with \(\Delta\) and varying  $(\lambda(\alpha)\) with $\alpha = 1$  in the range \([\lambda^s(\alpha),  \lambda^{s}(\alpha)+5.0]\), we observe that the eigenenergies are always real in that range, thereby confirming the  sufficient condition for the detecting unbroken phase.


\section{Bipartite entanglement and Parity of the zero-temperature state: Non-Hermitian and Hermitian systems }
\label{sec:ent}

In this section, we compare the properties of the ground state  for the $\mathcal{RT}$-symmetric and the corresponding  Hermitian systems. We calculate the parity and nearest-neighbor bipartite entanglement. Although the former feature exhibits the similarities between these two systems, the entanglement shows the opposite nature.   

\subsection{Bipartite entanglement}

We know that the factorisation point in the Hermitian systems corresponds to a completely factorised ground state  of the form \(|\psi_1\rangle \otimes |\psi_2\rangle \ldots \otimes |\psi_N\rangle\) with vanishing entanglement in all bipartitions. Let us examine the trends of entanglement at the surface where the broken-to-unbroken transition occurs in the \(iATXY\) model. In particular,  we find that  when we replace $\gamma$ by $i\gamma$,  the exceptional point, \(\lambda_1^s\) is,  indeed not a factorisation surface.

We observe that the nearest-neighbor  entanglement,  $E_{12}$, of the reduced bipartite state obtained from the zero-temperature state  is nonvanishing at the exceptional surface given in (\ref{realEq}) as depicted in Fig. \ref{fig:NNent}.  Notice that due to the translational symmetry of the model, all two-party nearest-neighbor state is same, and hence, we calculate the logarithmic negativity \cite{neg4, log_neg} of  \(\rho_{12}\) which is obtained after tracing out all the parties except the first and the second parties. We also find that the ground state is degenerate and hence we compute the bipartite entanglement of the canonical equilibrium state, \(\rho^{\beta} = e^{-\beta H_{iATXY}}/\mbox{Tr} (e^{-\beta H_{iATXY}}) \), with a very high \(\beta = 1/ K_B T = 200 \), \(T\) being the temperature and \(k_B\) being the Boltzmann constant. We call it the  zero-temperature state.

It can be explained by considering a general two-site density matrix between spins $a$ and $b$ which can be described in Pauli basis, $ \sigma^{i=x,y,z} $, as 
\begin{widetext}
$\begin{aligned} \rho_{ab}\left(m_{i}, m_{i}^{\prime}, C_{ij}\right)=& \frac{1}{4}\left(I_{4}+\sum_{i=x, y, z}\left[m_{i}\left(\sigma^{i} \otimes I_{2}\right)+m_{i}^{\prime}\left(I_{2} \otimes \sigma^{i}\right)+C_{i i}\left(\sigma^{i} \otimes \sigma^{i}\right) + \sum_{i\ne j=x,y,z}C_{i j}\left(\sigma^{i} \otimes \sigma^{j}\right)\right]\right) \\, \end{aligned}$

where $m_i = Tr(\rho_{ij} \sigma^i)$ and $C_{ij} = Tr(\rho_{ij}\sigma^i \otimes \sigma^j)$. The matrix form of \(\rho_{ab}\) can be written as 

\begin{equation}
\rho_{ab} = \frac{1}{4}\left(\begin{array}{llll}
1+C_{zz}+m_z+m_z^\prime &C_{zx}+m_x-im_y &m_x^\prime-im_y^\prime-iC_{yz} &C_{xx}-iC{xy}-C_{yy} \\
C_{zx}+m_x+im_y &1-C_{zz}-m_z+m_z^\prime & C_{xx}+iC{xy}+C_{yy} & m_x^\prime-im_y^\prime+iC_{yz}  \\
m_x^\prime+im_y^\prime+iC_{yz}& C_{xx}-iC{xy}+C_{yy}&1-C_{zz}+m_z-m_z^\prime &-C_{zx}+m_x-im_y\\
C_{xx}+iC{xy}-C_{yy} &m_x^\prime+im_y^\prime-iC_{yz} & -C_{zx}+m_x+im_y & 1+C_{zz}-m_z-m_z^\prime 
\end{array}\right)
\end{equation}

After taking the partial transposition over the party $a$, the matrix looks like

\begin{equation}
\rho_{ab}^{T_b}=\frac{1}{4}\left(\begin{array}{llll}
1+C_{zz}+m_z+m_z^\prime &C_{zx}+m_x+im_y  &m_x^\prime-im_y^\prime-iC_{yz} &C_{xx}+iC{xy}+C_{yy}  \\
C_{zx}+m_x-im_y &1-C_{zz}-m_z+m_z^\prime & C_{xx}-iC{xy}-C_{yy}  & m_x^\prime-im_y^\prime+iC_{yz}  \\
m_x^\prime+im_y^\prime+iC_{yz}& C_{xx}+iC{xy}-C_{yy} &1-C_{zz}+m_z-m_z^\prime & -C_{zx}+m_x+im_y\\
C_{xx}-iC{xy}+C_{yy} &m_x^\prime+im_y^\prime-iC_{yz} & -C_{zx}+m_x-im_y & 1+C_{zz}-m_z-m_z^\prime

\end{array}\right)
\end{equation}.

\end{widetext}
From the above form, it is clear that $\rho_{ab} = \rho_{ab}^{T_b}$ when  $\{m_y, C_{xy}, C_{yy}\} = 0 $ and hence, entanglement is vanishing.  In the case of the Hermitian system, since the Hamiltonian is real, the ground state should contain only real numbers which leads to  $\{m_y\) ,\( C_{xy}\}$ being vanishing. Moreover, at the factorization point, $C_{yy}$ also vanishes for the ground state. On the other hand,  in case of $\mathcal{RT}$-symmetric Hamiltonian, containing imaginary terms,   $\{m_y, C_{xy}, C_{yy}\}\ne 0$ for the ground state at the exceptional point. This  leads to a nonvanishing entanglement even at the exceptional point. 
The results possibly indicate that introducing \(\mathcal{RT}\) symmetry in the system is another way to generate entanglement in the factorization surfaces (cf. \cite{Fact1}). 

\subsection{Parity}

As defined above, the parity operator $\xi$ commutes with both the Hermitian and non-Hermitian Hamiltonians, i.e., $[\xi,H]=0$. It leads to their eigenstates having a definite parity, $\mu = \pm 1$. In the case of the Hermitian system, the parity of the ground state changes from negative to positive at the factorization point.  The similar  change of parity occurs for the  $\mathcal{RT}$-Symmetric Hamiltonian at the exceptional point.

\section{Conclusion} 
\label{sec:conclu}

We  found  that the factorization points of Hermitian quantum spin models dictate the exceptional points for the corresponding  rotation-time ($\mathcal{RT}$)-symmetric non-Hermitian Hamiltonians. As a demonstration, we analytically proved that the exceptional points of the non-Hermitian $XY$ model with uniform and alternating transverse magnetic fields ( the \(iATXY\) model) match the expression for the factorization surface of the nearest neighbor $ATXY$ model when the anisotropy parameter of the $ATXY$ model is replaced by the imaginary one. Following this prescription, we were able to predict and numerically verify the exceptional points of  the nearest-neighbor $iXYZ$ model. The other possible $\mathcal{RT}$-symmetric models considered here are long-range models, whose exceptional points are hard to find numerically. Hence we provided  a sufficient condition for obtaining the real energy spectrum  using  the factorization surface of the corresponding Hermitian model. Specifically, we observed that as long as the parameters of the non-Hermitian long-range \(iATXY\) and \(iXYZ\) models are above the factorization-like surfaces, the energy spectrum  are always  real. Moreover, at the exceptional points, we computed the bipartite entanglement of the nearest-neighbor two-site reduced state obtained from the ground state   and showed that it is nonvanishing although entanglement vanishes at the factorization surface of the Hermitian counterpart.

Quantum spin models with higher dimensional lattices as well as with long-range interactions can be studied only by using approximate methods or by numerical techniques. On the other hand, finding real or complex eigenvalues of the non-Hermitian spin models requires careful analysis of the entire  energy spectrum which is a difficult numerical task as also mentioned in Ref. \cite{PTSymmetry_numerical_arnoldi}. Therefore, the method proposed here to uncover the unbroken phase of the non-Hermitian models could  be a useful mechanism to bypass the extensive numerical search.  
 
 \section*{Acknowledgement}
We thank Shiladitya Mal for useful discussions. We also acknowledge the support from the Interdisciplinary Cyber Physical Systems (ICPS) program of the Department of Science and Technology (DST), India, Grant No.: DST/ICPS/QuST/Theme- 1/2019/23. The authors acknowledge computations performed at the cluster computing facility of the Harish-Chandra Research Institute, Allahabad, India. Some numerical results were obtained using the Quantum Information and Computation library (QIClib).  

\bibliographystyle{apsrev4-2}
\bibliography{bib}

\begin{thebibliography}{64}
\bibitem [{log()}]{log_neg}%
  \BibitemOpen
  \bibinfo {note} {Entanglement of a bipartite state, \(\rho_{ij}\), can be
  measured by logarithmic negativity, which is defined as follows:
  \begin{equation*} E(\rho_{ij}) = \log_2 [2 {\cal N}(\rho_{ij}) + 1],
  \end{equation*} where $\mathcal{N}$ quantifies the sum of absolute values of
  all negative eigenvalues of partially transposed matrix, denoted by
  $\rho_{ij}^{T_i}$.}\BibitemShut {Stop}%
\end{thebibliography}
\end{document}